\documentclass[runningheads]{svmult}

\usepackage{makeidx}   % allows index generation
\usepackage{graphicx}  % standard LaTeX graphics tool
\usepackage{subeqnar}  % subnumbers individual equations
\usepackage{multicol}  % used for the two-column index
\usepackage{physprbb}  % modified textarea for proceedings,
\makeindex             % used for the subject index
\begin{document}

\title*{Stochastic Dynamics of a Vortex Loop.\protect\newline
Thermal Equilibrium}
\toctitle{Stochastic Dynamics of a Vortex Loop.\protect\newline
Thermal Equilibrium}
% allows explicit linebreak for the table of content
%
%
\titlerunning{Stochastic Dynamics of a Vortex}
% allows abbreviation of title, if the full title is too long
% to fit in the running head
%
\author{S.K. Nemirovskii\inst{1} \and L.P. Kondaurova\inst{1}
\and M.Tsubota\inst{2}}
\authorrunning{S.K. Nemirovskii et al.}

\institute{Institute for Thermophysics, Lavrentyeva,1, 630090
Novosibirsk, Russia \and Department of Physics, Osaka City
University, Osaka, Japan}
 \maketitle

\begin{abstract}
We study stochastic behavior of a single vortex loop appeared in
imperfect Bose gas. Dynamics of Bose-condensate is supposed to
obey Gross-Pitaevskii equation with additional noise satisfying
fluctuation-dissipation relation. The corresponding Fokker-Planck
equation for probability functional has a solution
$\mathcal{P}(\{{\psi }(\mathbf{r})\})=\mathcal{N}\exp (-H\left\{ {
\psi }(\mathbf{r)}\right\} /T),$ where $H\left\{ {\psi
}(\mathbf{r})\right\} $ is a Ginzburg-Landau free energy.
Considering a vortex filaments as a topological defects of the
field ${\psi }(\mathbf{r})$ we derive a Langevin-type equation of
motion of the line with correspondingly transformed stirring
force. The respective Fokker-Planck equation for probability
functional $\mathcal{P}(\{\mathbf{s}(\xi )\})$ in vortex loop
configuration space is shown to have a solution of the form
$\mathcal{P}(\{ \mathbf{s}(\xi )\})=\mathcal{N}\exp (-H\left\{
\mathbf{s}\right\} /T),$ where $\mathcal{N}$ is a normalizing
factor and $H\left\{ \mathbf{s}\right\} $ is energy of vortex line
configurations. In other words a thermal equilibrium of
Bose-condensate results in a thermal equilibrium of vortex loops
appeared in Bose-condensate. Some consequences of that fact and
possible violations are discussed.
\end{abstract}

\section{Introduction and scientific background}

\indent Quantized vortices appeared in quantum fluids have been an
object of intensive study for many years (see for review and
bibliography the famous book by Donnelly\cite{Don_book}). The
greatest success in investigations of dynamics of quantized
vortices has been achieved in relatively simple cases such as
vortices in rotating helium (where they form a vortex array
orientated along an axis of rotation) or vortex rings. However
these simple cases are rather exception than a rule. Due to
extremely involved dynamics initially straight lines or rings
evolve to form highly entangled chaotic structure. Thus a
necessity of use of statistic methods to describe chaotic vortex
loop configurations arises. A most tempting way is to treat
''gas'' vortices as a kind of excitation and to use thermodynamic
methods. One of first examples of that way was an use of the
Landau criterium for critical velocity where vortex energy and
momentum were applied to relation having pure thermodynamic sense.
More extended examples would be the famous Kosterlitz-Thouless
description of 2D vortices or its 3D variant intensively
elaborated currently (for review and bibliography see
e.g.\cite{Williams}).
\\ \indent    In the examples above and in many other it is assumed that
chaotic vortex loop configurations are in a thermal equilibrium
and their statistics obeys the Gibbs distribution . That
supposition is based on fundamental physical principles and can be
justified in a standard way considering vortex lops as a subsystem
submerged into a thermostat and exchanging energy with the latter.
The role of thermostat in the case of vortices is played by the
other excitations (phonons and rotons) of an underlying physical
field. In the case of Bose- Einstein condensate (BEC), which we
consider in this paper, that field is an order parameter $\psi (
\mathbf{r,}t)$. Vortex lines are just the crossings of surfaces
where both real and imaginary part of $\psi (\mathbf{r,}t)$
vanish. Excitations of order parameter (phonons and rotons)
interact with vortices driving the latter to statistic
distribution which in accordance with general principles should be
the Gibbs distribution.\\ \indent It is well known however that
the Gibbs distribution can be alternatively obtained in the frame
of some reduced model like kinetic equations or Fokker-Planck
equation. That way of course is not of such great generality as a
principle of maximum entropy, but instead it allows us to clarify
the mechanisms of how the Gibbs distribution is established and to
discuss possible deviations and violations of equilibrium state.
We choose that way to examine how a thermal equilibrium of chaotic
vortex loops follows from a thermal equilibrium of BEC.

\section{Langevin equation}

We perform our consideration  on the basis of the Gross-Pitaevskii
model\cite {Pit58},\cite{Gross61}. To describe stochastic behavior
of BEC let us add to the right-hand side of the Gross-Pitaevskii
equation the random stirring force $\zeta (\mathbf{x},t).$
\begin{equation}
\frac \hbar m\frac{\partial \psi }{\partial t}=-\left( \Lambda +i\right)
\frac{\delta H(\psi )}{\delta \psi ^{*}}+\zeta (\mathbf{x},t).
\label{Canonical1}
\end{equation}
Here $H\left\{ \psi \right\} $-is the Ginzburg-Landau free energy functional

\begin{equation}
H\left\{ \psi \right\} =\int d^3x\left[ \frac{\hbar ^2}{2m^2}\left| \nabla
\psi \right| ^2-\frac \mu m\left| \psi \right| ^2+\frac{V_0}{2m}\left| \psi
\right| ^4\right].  \label{H(psi)}
\end{equation}
The thermal noise obeys the following fluctuation-dissipation theorem
\begin{equation}
\left\langle \zeta (\mathbf{x}_1,t_1)\zeta ^{*}(\mathbf{x}
_2,t_2)\right\rangle ~=\frac{2k_BT\Lambda }{(m/\hbar )}~\delta
(\mathbf{x}_1- \mathbf{x}_2)~\delta (t_1-t_2)~.  \label{fdt2}
\end{equation}
The stochastic problem introduced by
(\ref{Canonical1})-(\ref{fdt2}) has a solution describing thermal
equilibrium, where probability of some configuration of $\psi
$-filed is proportional to $\exp (-H\left\{ {\psi }(
\mathbf{r)}\right\} /T)$. This readily follows from the
correspondent Fokker-Planck equation\cite{Hohenberg}.

Let's go on to the problem of obtaining of the vortex dynamics appearing in
BEC, whose own dynamics obeys relations (\ref{Canonical1})-(\ref{fdt2}).
That problem, as well as more general problems of that kind has been
considered many times by many methods\cite{rem1} starting from pioneering
papers by Pitaevskii\cite{Pit61} and Fetter\cite{Fetter66}. We develop one
more method, neither too rigorous nor principally new but convenient for
purpose of this paper. Details of that method will be published elsewhere,
here we briefly describe it to an extent to comprehend the put goal.

As any motion of topological defects the vortex line dynamics is
determined by the one of the underlying field theory. On the other
hand if one ignores a presence of other excitations except of the
vortices one can say that all motion of BEC is determined by the
one of quantum vortices. Thus there is mutual correspondence and
the order parameter $\psi (\mathbf{x},t)$ can be considered as
some functional of a whole vortex loop configuration $\psi (
\mathbf{x\mid \{s(}\xi ,t)\}).$ Temporal dependance of field $\psi
$ is connected with motion of lines and its rate of change (at
some point $ \mathbf{x}$) is expressed by the following chain
rule:
\begin{equation}
\frac{\partial \psi (\mathbf{x},t)}{\partial t}=\int\limits_\Gamma
\frac{ \delta \psi (\mathbf{x\mid \{s(}\xi ,t)\})}{\delta
\mathbf{s(}\xi ^{\prime },t)}\frac{\partial \mathbf{s(}\xi
^{\prime },t)}{\partial t}d\xi ^{\prime }. \label{dpsi/dt}
\end{equation}
Vortices in quantum fluids are very slender tubes (except of
vicinity of phase-transition) and their dynamics is analogous to
the one of the strings. That implies that we have to aim our
efforts at integrating out radial degrees of freedom. It can be
reached by the following procedure. Let us further multiply
equation(\ref{Canonical1}) by $\frac 1{-\Lambda +i}$ $ \delta \psi
^{*}/\delta \mathbf{s(}\xi _0,t),$ where $\xi _0$ some chosen
point on the curve. Combining the result obtained with the complex
conjugate and integrating over whole space we have
\begin{eqnarray}
&&\frac \hbar m\int d^3\mathbf{x}\left( \frac{\Lambda -i}{\Lambda
^2+1}\frac{
\partial \psi }{\partial t}\frac{\delta \psi ^{*}}{\delta \mathbf{s(}\xi
_0,t)}+\frac{\Lambda +i}{\Lambda ^2+1}\frac{\partial \psi
^{*}}{\partial t} \frac{\delta \psi }{\delta \mathbf{s(}\xi
_0,t)}\right) =  \nonumber \\ &&-\int d^3\mathbf{x}\left(
\frac{\delta H(\psi ,\psi ^{*})}{\delta \psi ^{*} }\frac{\delta
\psi ^{*}}{\delta \mathbf{s(}\xi _0,t)}+\frac{\delta H(\psi ,\psi
^{*})}{\delta \psi ^{*}}\frac{\delta \psi }{\delta \mathbf{s(}\xi
_0,t) }\right) +~  \label{Can3} \\ &&\int d^3\mathbf{x}\left(
\frac{\Lambda -i}{\Lambda ^2+1}\zeta (\mathbf{x} ,t)\frac{\delta
\psi ^{*}}{\delta \mathbf{s(}\xi _0,t)}+\frac{\Lambda +i}{ \Lambda
^2+1}\zeta ^{*}(\mathbf{x},t)\frac{\delta \psi }{\delta
\mathbf{s(} \xi _0,t)}\right) .  \nonumber
\end{eqnarray}
The first integral in the right-hand side of (\ref{Can3}) expresses a chain
rule for functional derivative $\delta H(\mathbf{s})/\delta \mathbf{s(}\xi
_0,t)$ where $H(\mathbf{s})$ is the energy of moving BEC expressed via
vortex line position. Consequently considering them to be very slender tubes
(which is justified when the radius of the curvature $R$ is much larger of
the core size $r_0$) and neglecting an energy associated with the core, the
quantity $H(\mathbf{s})$ is just kinetic energy of the superfluid flow
created by vortices (see e.g. \cite{Bat},\cite{Saf})

\begin{equation}
H(\mathbf{s})=\frac{{\rho }_s{\stackrel{\symbol{126}}{\kappa
}}^2}{8\pi } \int\limits_\Gamma \int\limits_{\Gamma ^{^{\prime
}}}\frac{\mathbf{s} ^{\prime }(\xi )\mathbf{s}^{\prime }(\xi
)}{|\mathbf{s}(\xi )-\mathbf{s}(\xi ^{\prime })|}d\xi d\xi
^{\prime }.  \label{H(s)}
\end{equation}
Here $\mathbf{s}^{\prime }(\xi )$ is tangent vector, double integration is
performed along a whole line, {$\stackrel{\symbol{126}}{\kappa }$} is a
quantum of circulation equal $2\pi \hbar /m$. Calculation of functional
derivative $\delta H(\mathbf{s})/\delta \mathbf{s(}\xi _0,t)$ is
straightforward and leads to result
\begin{equation}
\frac{\delta H(\mathbf{s)}}{\delta \mathbf{s(}\xi _0,t)}={\rho
}_s{\stackrel{ \symbol{126}}{\kappa }}\mathbf{s}^{\prime }(\xi
_0)\times \mathbf{B(}\xi _0). \label{dH/ds}
\end{equation}
Quantity $\mathbf{B(}\xi _0)$ is the velocity of the line element
$\stackrel{ \cdot }{\mathbf{s}}\mathbf{(}{\ \xi }_0)$ expressed by
well known Biot-Savart law. The terms in the left-hand side of
equation (\ref{Can3}) (first line) can be evaluated in general
form by observing that the major contribution into integrals
appears from vicinity of the vortex filament (see
e.g.\cite{Pismen}). Thus to evaluate integral we replace $\psi
(\mathbf{ x\mid s(}\xi ,t))$ by $\psi _v(\mathbf{x}_{\perp })=\psi
_v(\mathbf{s(}\xi _{cl},t)-\mathbf{x})$ where $\psi _v$ is well
studied 2D vortex and integration over $d^3\mathbf{x}$ by
$d^2\mathbf{x}_{\perp }d\xi _{cl}$. Functional derivative $\delta
\psi ^{*}/\delta \mathbf{s(}\xi ^{\prime },t)$ should be evaluated
by a following rule: $\delta \psi ^{*}/\delta \mathbf{s(} \xi
^{\prime },t)=\nabla _{\perp }\psi _v(\mathbf{x}_{\perp })\delta
(\xi ^{\prime }-\xi _{cl}).$ Here $\xi _{cl}$ is the label of
point of the line closest to point $\mathbf{x}$. Using the said
above and calculating integrals of squared gradients of $\psi
_v(\mathbf{x}_{\perp })$ we conclude that left-hand side of
equation (\ref{Can3}) transforms into

\begin{equation}
\frac \hbar m\frac{2\pi \rho _s}{\Lambda ^2+1}\stackrel{\cdot
}{\mathbf{s}} \mathbf{(}{\ \xi }_0)\times \mathbf{s}^{\prime }(\xi
_0)+\frac \hbar m\frac{ 2\pi \rho _s\sigma \Lambda }{\Lambda
^2+1}\stackrel{\cdot }{\mathbf{s}} \mathbf{(}{\ \xi }_0).
\label{s_dot_fin}
\end{equation}

Let us now discuss the rest terms of equation (\ref{Can3})
including random force $\zeta (\mathbf{x},t)$ (the third line).
Consequently considering that the all motion of BEC is connected
to motion of line, we have to consider Langevin force $\zeta
(\mathbf{x},t)$ as some secondary quantity stemming from random
displacements of filaments. Connection between displacements
(random) of filaments $\delta \mathbf{s}$ and deviations (random)
of $\delta \psi (\mathbf{x},t)$ may be written in form similar to
(\ref{dpsi/dt}) with formal substitution $\partial \psi
(\mathbf{x},t)/\partial t\rightarrow \delta \psi (\mathbf{x},t)$
and $\stackrel{\cdot }{\mathbf{s}}\rightarrow \delta \mathbf{s}$.
Taking into account that $\delta \psi (\mathbf{x} ,t)=\zeta
(\mathbf{x},t)\delta t$, and $\delta \mathbf{s=\zeta (}\xi
,t)\delta t$ we conclude that quantities $\zeta (\mathbf{x},t)$
and $\mathbf{ \zeta (}\xi ,t)$ are connected to each other by a
chain rule
\begin{equation}
\zeta (\mathbf{x},t)=\int\limits_\Gamma \frac{\delta \psi
(\mathbf{x\mid \{s( }\xi ,t)\})}{\delta \mathbf{s(}\xi ^{\prime
},t)}\mathbf{\zeta (}\xi ^{\prime },t)d\xi ^{\prime }.
\label{dz-dz}
\end{equation}

That implies that to take into consideration random displacements
of line we have to change the last term in equation (\ref{Can3})
by the one similar to ( \ref{s_dot_fin}) with substitution
$\stackrel{\cdot }{\mathbf{s}}\mathbf{(}{ \ \xi }_0)\rightarrow
\mathbf{\zeta (}\xi _0,t)$. Gathering all terms we obtain a vector
equation, which can be resolved up to tangential velocity $
\stackrel{\cdot }{\mathbf{s}_{\parallel }}\mathbf{(}{\ \xi }_0)$
along the curve. The latter does not have any physical meaning and
can be removed by suitable parameterization of the label variable
$\xi $. Solving that vector equation we arrive at
\begin{equation}
\stackrel{\cdot }{\mathbf{s}}\mathbf{(}{\ \xi }_0)=\frac{1+\Lambda
^2}{ 1+\Lambda ^2\sigma ^2}\mathbf{B(}\xi _0)+\frac{(1+\Lambda
^2)\Lambda \sigma }{1+\Lambda ^2\sigma ^2}\mathbf{s}^{\prime }(\xi
_0)\times \mathbf{B(}\xi _0)+(m/\hbar )\mathbf{\zeta (}\xi _0,t).
\label{master}
\end{equation}
Equation (\ref{master}) describes motion of vortex line in terms
of the line itself. It is remarkable fact (not obvious in advance
that noise $\mathbf{ \zeta (}\xi _0,t)$ acting on line is also
additive (does not depend on line variables).

The last effort we have to do is to ascertain both the statistic
properties of noise $\mathbf{\zeta (}\xi _0,t)$ and its intensity.
Shortly, it can be done by comparison of equation(\ref{s_dot_fin})
with substitution $\stackrel{ \cdot }{\mathbf{s}}\mathbf{(}{\ \xi
}_0)\rightarrow \mathbf{\zeta (}\xi _0,t) $ with the last term of
equation (\ref{Can3}). Clearly the former appeared as result of
transformation of the latter. Equating them and taking the scalar
productions of both parts of the resulting relation we arrive at
\begin{equation}
\left\langle \mathbf{\zeta }_{\eta _1}\mathbf{(}\xi
_1,t_1)\mathbf{\zeta } _{\eta _2}\mathbf{(}\xi
_2,t_2)\right\rangle ~=\frac{k_BT}{\rho _s\pi (\hbar
/m)}\frac{\left( \Lambda ^2+1\right) \Lambda \sigma }{1+\Lambda
^2\sigma ^2} ~\delta (\xi _1\mathbf{-}\xi _2)~\delta
(t_1-t_2)\delta _{\eta _1,\eta _2}~. \label{fdt_line}
\end{equation}
Here $\mathbf{\zeta }_{\eta _1}$ and\textbf{\ }$\mathbf{\zeta
}_{\eta _2}$ \textbf{\ }are components of random velocities in
$\eta _1,\eta _2$ directions lying in the plain normal to the
line.

Thus starting from dynamics of BEC ( equation (\ref{Canonical1}))
with the fluctuation-dissipation theorem (\ref{fdt2}) we derive
equation (\ref{master} ) describing motion of vortex line in terms
of line itself with the additive noise obeying the
fluctuation-dissipation theorem (\ref{fdt_line}). These relations
complete a stochastic problem of quantized vortex dynamics under
thermal noise stemming from the one stirring the underlying field.
In the next section we demonstrate that this problem has an
equilibrium solution given by Gibbs distribution $\exp (-H\left\{
\mathbf{s}\right\} /k_BT),$ where $H\left\{ \mathbf{s}\right\} $-
functional of energy due to vortex loop (equation (\ref{H(s)}))
and $T$ -temperature of Bose-condensate.

\section{ Fokker-Planck equation}

To show it we, first, derive the Fokker-Planck equation corresponding to
Langevin type dynamics obeyed (\ref{master}) and (\ref{fdt_line}). Let us
introduce probability distribution functional (PDF)
\begin{equation}
\mathcal{P}(\{\mathbf{s}(\xi )\},t)=\left\langle \delta \left(
\mathbf{s} (\xi )-\mathbf{s}(\xi ,t)\right) \right\rangle .
\label{pdf}
\end{equation}
Here $\delta $ is delta functional in space of vortex loop configurations.
Averaging is fulfilled over ensemble of random force. The Fokker-Planck
equation can be derived in standard way (see e.g. \cite{Zinn-Justin96})
\begin{eqnarray}
&&\frac{\partial \mathcal{P}}{\partial t}+\int d\xi \frac \delta
{\delta \mathbf{s}(\xi )}\left[ \frac{1+\Lambda ^2}{1+\Lambda
^2\sigma ^2}\mathbf{B(} \xi )+\frac{(1+\Lambda ^2)\Lambda \sigma
}{1+\Lambda ^2\sigma ^2}\mathbf{s} ^{\prime }(\xi )\times
\mathbf{B(}\xi )\right] \mathcal{P}+  \label{FP2} \\ &&\int \int
d\xi d\xi ^{\prime }\frac{k_BT}{2\pi \rho _s(\hbar /m)^2}\frac{
\left( \Lambda ^2+1\right) \Lambda \sigma }{1+\Lambda ^2\sigma
^2}~\delta (\xi \mathbf{-}\xi ^{\prime })~\delta (t_1-t_2)\delta
_{\eta _{1} ,\eta _2}~\frac \delta {\delta \mathbf{s}(\xi )}\frac
\delta {\delta \mathbf{ s}(\xi ^{\prime })}\mathcal{P}=0 \nonumber
\end{eqnarray}
Equation (\ref{FP2} ) possesses the equilibrium solution in form
of the Gibbs distribution $\mathcal{P}(\{\mathbf{s}(\xi
)\})=\mathcal{N}\exp (-H\left\{ \mathbf{s}\right\} /T),$ where
$\mathcal{N}$ is a normalizing factor. Let us show that the first
integral term vanishes identically for that solution. To do it we
exploit relation (\ref{dH/ds}) and parametrization of label
variable $\xi $ in which velocity  $\stackrel{\cdot
}{\mathbf{s}}({\xi }_0)$ is normal to the line. Using a tensor
notation we rewrite the first term in integrand in form (we omit
the coefficient and factor $\exp (-H\left\{ \mathbf{s} \right\}
/k_BT)$)
\[
 \epsilon ^{\alpha \beta \gamma }\left\{ \frac{\delta
\mathbf{s} _\beta ^{\prime }(\xi )}{\delta \mathbf{s}_\alpha (\xi
)}\frac{\delta H( \mathbf{s)}}{\delta \mathbf{s}_\gamma
\mathbf{(}\xi ,t)}+
  \mathbf{s}_\beta ^{\prime }(\xi )
\frac{\delta^{2} H(\mathbf{s)}}{\delta \mathbf{s}_\gamma
\mathbf{(}\xi ,t){\delta \mathbf{s}_\alpha (\xi )}}
+\mathbf{s}_\beta ^{\prime }(\xi )\frac{\delta
H(\mathbf{s)}}{\delta \mathbf{s}_\gamma \mathbf{(}\xi
,t)}\frac{\delta H(\mathbf{s)}}{\delta \mathbf{s}_\alpha
\mathbf{(}\xi ,t)}\right\}
\]
The functional derivative $\delta \mathbf{s}_\beta ^{\prime }(\xi
)/\delta \mathbf{s}_\alpha (\xi )\propto \delta _{\beta \alpha }$
therefore all terms vanish due to symmetry. Thus the reversible
term gives no contribution to flux of probability (in the
configuration space) equation (\ref{FP2}), one says it is
divergence free. Furthermore exploiting again relation \ref{dH/ds}
) one convinces himself that second (dissipative) term in
(\ref{FP2}) and third (due to stirring force) term exactly
compensate each other (locally) as it should be in the thermal
equilibrium.

\section{Possible violation of thermal equilibrium.}

Thus we have proved that the thermal equilibrium of BEC results in
the thermal equilibrium of vortex loop. We are now in position to
discuss how it can be destroyed. Analyzing the proof one can see
that the following steps were crucial. 1. Additive white noise
$\zeta (\mathbf{x},t)$ acting on field $\psi (\mathbf{x},t)$ is
transformed into additive white noise\textbf{\ }$ f(\xi ,t)$
acting on vortex line position $\mathbf{s(}\xi \mathbf{,}t\mathbf{
).}$ 2. Intensity of noise $f(\xi ,t)$ expressed by
(\ref{fdt_line}) is that it locally compensates dissipative flux
of probability distribution functional in the Fokker-Planck
equation (\ref{FP2} ).

That observation points out how the thermal equilibrium in space
of vortex loops can be destroyed for real vortex tangles appeared
e.g. in counterflowing HeII or formed in quenched superfluids.
Being a macroscopical objects vortex loops inevitably undergo
large scale perturbations generated e.g. by nonuniform flow or by
action of other vortex loops, randomly placed with respect to the
studied loop. One more essential source of large scale
perturbations might be long wave instabilities of vortex filament
motion. That type of random action drastically differs from
$\delta $-correlated in $ \xi $ space thermal noise considered
above. It obviously cannot compensate dissipative flux of
probability, which is proportional curvature and acts accordingly
in small scales. Instead the following scenario seems to be
realized\cite{npp91}. Due to nonlinear character of the equation
of motion the large-scale perturbations on an initially smooth
filament interact creating higher harmonics. They in turn generate
harmonics with larger $ \kappa $ , where $\kappa $ is
one-dimensional wave vector arisen in 1D Fourier transform (with
respect to label variable $\xi $) of quantity $ \mathbf{s}(\xi
,t)$. One can say that an additional curvature created by large
scale stirring force propagates in region of small scales. In real
space that corresponds to entangling of vortex loop and creation
of vortex line segments with large local curvature. Then
dissipative processes come into play, their role now is that they
remove from the system harmonics with very large wave vectors
$\kappa .\,$ So the Kolmogorov cascade-like solution with a flux
of a curvature in space of $\dot{\kappa}$ is established and, as a
result, the stochastic distribution is far from equilibrium. That
scenario is quite similar to the one which is realized in
classical turbulence or in the so called weak (wave) turbulence. A
difference is that if in case of the wave and classical
turbulence) an exchange an energy between harmonics is realized,
whereas in our case there is exchange a curvature $\left\langle
\kappa ^4\mathbf{s}(\kappa )\mathbf{s}(-\kappa )\right\rangle $.

The scenario described above is especially relevant in the so
called low-temperature superfluid turbulence case, when the normal
component is very small and the usual (considered here)
dissipation monotonically increasing with the curvature is absent.
Some recent experiments and numerical simulations\cite{TON00} show
that vortex tangle in HeII decays at extremely low temperature
(about 1 mK), where dissipation due to normal component is
negligibly small. Obviously some other strong mechanisms of
dissipation must take place. It can be e.g. emission of phonons
and rotons from speedy moving parts of line or just collapse of
''hairpin'' segments of the filament. That mechanisms are
concentrated on very small scales (of order of the vortex core
size) or equivalently in region of very large wave numbers $\kappa
$. Therefore regions of the pumping and the sink of an additional
curvature are greatly remote in $\kappa $ space and a system must
be essentially nonequilibrium.

One more reason of a violation of thermal equilibrium might be a
reconnection of lines. In real vortex tangle consisting of many
loops the vortex filaments undergo frequent collisions and
reconnections. Just after reconnection there appear kinks on the
curves disappearing later on. From mathematical point of view a
kink on the curve can be described as a discontinuity of tangent
vector $\mathbf{s}^{\prime }(\xi ,t)$ which has Fourier transform
of type $\mathbf{s}^{\prime }(\kappa )\propto $ $\kappa ^{-1}$.
Thus the reconnection processes supply the selected curve with
discontinuities having a spectrum $\mathbf{s}(\kappa )\propto
\kappa ^{-2}$ . Taking into account a random nature of vortex line
collision the reconnection processes can be modelled, in some
measure as a random stirring of filament with spectrum of type
$\left\langle \mathbf{\zeta }(\kappa ) \mathbf{\zeta }(-\kappa
)\right\rangle \propto $ $\kappa ^{-4}$. Remember now that the
establishing of a thermal equilibrium requires that the random
force correlation function
  is proportional $\propto \delta (\xi _1\mathbf{-}\xi
_2)$ and, consequently the spectrum does not depend on wave number
$\kappa $, $\left\langle \mathbf{ \zeta }(\kappa )\mathbf{\zeta
}(-\kappa )\right\rangle =const$ . Therefore the colored noise
$\left\langle \mathbf{\zeta }(\kappa )\mathbf{\zeta } (-\kappa
)\right\rangle \propto $ $\kappa ^{-4}$ coming from reconnection
processes can also lead to the Kolmogorov type nonequilibrium
state. To clarify which of mechanisms forming nonequilibrium state
prevails in real vortex tangle one has to investigate much more
involved problem.

This work was partly supported under grant N 99-02-16942 from Russian
Foundation of Basic research.

%INDEX%%%%%%%%%%%%%%%%%%%%%%%%%%%%%%%%%%%%%%%%%%%%%%%%%%%%%%%%%%%%%%
% Please check with the editor of your book whether he plans to
% include a "mutual" subject index - if so, please code your entries
% in the standard syntax. For your own purposes you may print your
% "personal" index by using the following commands:
%
%\clearpage
%\addcontentsline{toc}{section}{Index}
%\flushbottom
%\printindex
%%%%%%%%%%%%%%%%%%%%%%%%%%%%%%%%%%%%%%%%%%%%%%%%%%%%%%%%%%%%%%%%%%%%%

\end{document}